\documentclass[review,5p,twocolumn,10pt]{elsarticle}

\usepackage[colorlinks,citecolor=red]{hyperref}  
\usepackage{threeparttable}  
\usepackage{booktabs}  

\usepackage{array}  

\usepackage[american]{babel}
\usepackage{microtype}
\usepackage{color}

\usepackage{graphicx}
\usepackage{caption}
\usepackage{graphicx}
\usepackage{dcolumn}
\usepackage{bm}
\usepackage{latexsym}
\usepackage{amsmath}
\usepackage{amsmath,amssymb,bm}
\usepackage{graphicx,color}
\usepackage{hyperref}
\usepackage{epstopdf}
\usepackage{mathrsfs}
\usepackage{float}
\usepackage{subfigure}
\usepackage{braket}

\newcommand{\be}{\begin{equation}}
\newcommand{\ee}{\end{equation}}

\usepackage{amssymb}
\usepackage{amsthm}
\usepackage{amsmath}

\usepackage{lineno}
\journal{The Journal of Chemical Physics}

\begin{document}


\begin{frontmatter}



\title{\textbf{ The Impact of Geometric Blockade on Thermoelectric Transport in Triangular Triple Quantum Dots  }}


\author{Shuo Dong $^{a}$}

\author{Yiming Liu $^{a}$}

\author{Junqing Li $^{a}$}

\author{Jianhua Wei $^{a*}$}
\ead{corresponding author: wjh@ruc.edu.cn}


\address[1]{School of Physics, The Renmin University of China, Beijing, 100876 China}

\begin{abstract}
We investigate the transport properties of a triangular triple quantum dot (TTQD) system connected with two reservoirs under linear response regime. By employing the hierarchical equations of motion(HEOM), we compute the thermopower and thermoelectric figure of merit. The impact of interaction scheme among three quantum dots on the thermopower is thoroughly analyzed, while the thermoelectric current and spectral function throughout this process are also elaborated.  Our results reveal that, under low-temperature conditions, the alleviation of the geometric blockade in the TTQD system leads to a significantly faster enhancement of the heat current compared to the electric current. This phenomenon consequently elevates the thermopower, resulting in a remarkably high thermoelectric figure of merit.
\end{abstract}

\begin{keyword}
logical entropy; quantum correlation measure; projection measurement; quantum system; bipartite quantum states
\end{keyword}
\end{frontmatter}

\section{Introduction}
Thermoelectric materials, which possess the ability to convert thermal energy into electrical energy, have garnered significant attention in recent years due to their potential applications in various fields, such as thermoelectric generators \cite{ref1}. The thermoelectric conversion efficiency of  a material is typically characterized using  thermoelectrical figure of merit $ZT$, while $ZT=S^{2} G T /\kappa$. Here $S$ is the thermopower, $ G $ is the electric conductance, and $\kappa $ is the total thermal conductance. Therefore,in order to get a high $ZT$, the thermopower and electric conductance are supposed to increase, and decrease the thermal conductance of the material. In conventional solid materials, the electrical and thermal conductance are intrinsically linked through the Wiedemann-Franz law\cite{ref2}, which indicates that we are unable to reduce the thermal conductance while enhancing the electrical conductance simultaneously. Meanwhile,the thermopower(S) follows the Mott relation\cite{ref3} $S_{\text {Mott }}=-\left.\frac{\pi^{2} k_{B}^{2} T}{3 e} \frac{\partial \ln G(E)}{\partial E}\right|_{E=E_{F}}$, $E_{F}$ is fermi energy. The Mott formula indicates that enhanced thermopower requires strong energy-dependent transport—specifically, a large logarithmic derivative of conductance at the Fermi energy.  This implies the need for sharp resonances or strong energy filtering in the density of states near the Fermi level. As a result, the thermoelectric efficiency of traditional solid materials is inherently limited. 
	
	However, in nanoscale devices such as quantum dots (QDs) and molecular junctions, the Mott relation may not be applicable, and the Wiedemann-Franz law could be infringed upon due to the influence of quantum confinement and Coulomb blockade effects \cite{ref4,ref5}, which opened up new avenues for enhancing thermoelectric efficiency\cite{ref6,ref7,ref8,ref9}.

	QDs systems exhibit a plethora of phenomena related to thermoelectric effects\cite{ref10,ref11,ref12} that deviate from those observed in conventional materials. Moreover, various quantum interference effects, such as Fano resonance \cite{ref13}, Dicke effect \cite{ref14}, and structures exhibiting Aharonov-Bohm (AB) effect \cite{ref15}, have been shown to substantially influence the thermoelectric properties of QDs systems. These unique characteristics of quantum dot systems offer exciting opportunities for the development of novel thermoelectric devices with improved performance and efficiency. Early theoretical studies on quantum-dot thermoelectrics also highlighted the role of discrete energy levels in enhancing the Seebeck coefficient. Beenakker and Staring, for example, explored thermopower in the Coulomb-blockade regime of a single quantum dot and predicted sharp peaks in S, which lead to substantial enhancements in $ZT$ at low temperatures \cite{ref21}. Theoretically, Humphrey et al. predicted that a single-dot heat engine could achieve $ZT$ approaching unity around 1-4 K by operating in the reversible limit, where the quantum dot energy level is symmetrically positioned between the chemical potentials of the hot and cold reservoirs \cite{ref25}. Experimentally, similar quantum dot thermoelectric effects have been observed by Josefsson et al. \cite{ref28}. Further investigations by Scheibner et al. into Kondo-correlated quantum dots revealed that many-body effects significantly modify both $S$ and $ \sigma $ at sub-Kelvin temperatures, showcasing the complex interplay between electron correlations and thermoelectric behavior \cite{ref26}. Gomez‑Silva et al. investigate a T-shaped double-QD system and find that quantum interference together with a Van‑Hove singularity in the leads’ density of states can strongly boost thermopower and the figure of merit, both in linear and nonlinear response regimes \cite{ref30}. Pirot et al. analyze a serial DQD with intra and inter dot Coulomb interactions, showing that these interactions open extra transport channels, which significantly enhance both heat and charge currents \cite{ref31}. Among these systems, the triangular triple quantum dot(TTQD) system emerges as a captivating subject of study. This configuration offers a rich parameter space and a distinct geometric structure, which allows for the manifestation of a diverse array of many-body phenomena in the strongly correlated regime\cite{ref16,ref17}. However, despite the potential of TTQD systems, investigations into their thermoelectric transport properties have been comparatively scarce in the current literature.
	
	According to our previous research\cite{ref18}, we have observed a geometric blocking effect in the TTQD structure. The geometric blocking effect arises when the TTQD system(the electrode part is not included) exhibits $C_{3v}$ symmetry, resulting in the generation of chiral flows due to spin-field coupling that occupy the transport channel and impede the transport current. However, by breaking the $C_{3v}$ symmetry, this blockade disappears, and the transport current is consequently restored.This blockade effect represents a distinctive and novel quantum phenomenon exclusive to the configuration of cyclic triple quantum dots. The introduction of a chirality operator by the non-coplanar arrangement of three spins endows the TTQD system with characteristics not observed in conventional quantum dot systems. Despite its significance, research delving into this effect is currently limited, with its impact on thermoelectric transport being particularly under-explored.

     We study the triangular three-quantum dot system by using hierarchical equations of motion. The hierarchical equations of motion (HEOM) method has emerged as a powerful non-perturbative approach for treating open quantum systems coupled to fermionic or bosonic reservoirs. Originally developed for studying quantum dissipation in bosonic baths, the method has been extended to fermionic environments and has proven particularly effective for quantum transport problems where strong system-lead coupling and electron-electron interactions must be treated on equal footing. Recent advances have significantly expanded the scope and efficiency of HEOM applications to quantum transport. Tanimura provided a comprehensive review of the theoretical foundations and numerical implementations of HEOM for both bosonic and fermionic systems, highlighting its advantages over perturbative master equation approaches in the strong-coupling regime \cite{ref29}. These developments establish HEOM as a reliable tool for exploring thermoelectric phenomena in multi-dot systems where both quantum coherence and many-body correlations play essential roles.

	In this work, we investigate the thermoelectric properties of a triangular triple quantum dot system, where two of the dots are individually connected to separate electrodes, as shown in figure 1(a). By employing the hierarchical equations of motion(HEOM) approach \cite{ref19,ref20}, we explore the evolution of the system's thermoelectric characteristics as the inter-dot coupling strength is varied, transitioning from a linear quantum dot configuration to a triangular arrangement with $C_{3v}$ symmetry, and subsequently to a triangular configuration with broken symmetry. Our analysis reveals that the geometric blockade, which arises in the symmetric triple quantum dot system, leads to a suppression of the thermoelectric figure of merit. Intriguingly, the deliberate breaking of the symmetry results in a substantial enhancement of the thermoelectric power.

\section{Theory and Methoed}
The architecture we examined comprises three quantum dots, each exhibiting mutual coupling, and in conjunction with two electrodes. 
	
	The total composite Hamiltonian is
	
	\begin{equation} 
		H_{T}=H_{S}+H_{B}+H_{S B},
	\end{equation}
	
	where $ H_{S} $ is the Hamiltonian for the three coupled dots.
	
	\begin{equation}
		H_{S}=\sum_{i \mu} \varepsilon_{i} \widehat{d}_{i \mu}^{\dagger} \widehat{d}_{i \mu}+U \sum_{i} \widehat{n}_{i \uparrow} \widehat{n}_{i \downarrow}+\sum_{i \neq j \mu} t_{i j} \widehat{d}_{i \mu}^{\dagger} \widehat{d}_{j \mu}.
	\end{equation} 
	
	The $d_{i \mu}^{\dagger}\left(d_{i \mu}\right)  $ in above fomula is creation(annihilation) operator for an electron with $\mu $ spin on the $i-th $ dot. And $H_{B}$ is the Hamiltonian for the leads.
	
	\begin{equation}
		H_{B}=\sum_{\alpha k \mu} \varepsilon_{\alpha k} c_{\alpha k \mu}^{\dagger} c_{\alpha k \mu} ,
	\end{equation}
	
	$ c_{\alpha k \mu}^{\dagger}\left(c_{\alpha k \mu}\right)$ is creation(annihilation) operator for electron of lead $ \alpha$ on the $ k -th $ state, and $\epsilon_{k \alpha}$  is the energy of an electron with wave vector  $k $ in the $ \alpha$  lead.
	
	The dot-electrode coupling Hamiltonian is $ H_{S B} $
	
	\begin{equation}
		H_{S B}=\sum_{\alpha k i \mu} V_{\alpha k i \mu} d_{i \mu}^{\dagger} c_{\alpha k \mu}+  H.c. ,
	\end{equation}
	
	with $V_{\alpha k i \mu \mu} $ being the tunnel matrix element between  $i -th$ impurity and electrons with $ k -th$ state on the  $\alpha$ -reservoir. For this paper, $ V_{\alpha k \mu} $ is the electron tunneling strength between two leads. The effect of electron reservoirs on QDs is taken into account through the hybridization functions, $ \Delta_{\mu v}(\omega) \equiv \Sigma_{\alpha} \Delta_{\alpha \mu \nu}(\omega)=\pi \Sigma_{\alpha k} V_{\alpha \mu k} V_{\alpha v k}^{*} \delta\left(\omega-\varepsilon_{\alpha k}\right) $, in the absence of applied chemical potentials. Generally, we adopt Lorentzian hybridization functions in the HEOM approach, that is, $ \Delta_{\mu v}(\omega)=\delta_{\mu v} \Delta W^{2} /\left(\omega^{2}+W^{2}\right) $, with $ \Delta=\Sigma_{\alpha} \Delta_{\alpha} $ being the overall dot-lead coupling strength and  W  the bandwidth of the electrodes.
	
	When a magnetic flux $ \phi $ is present within the triangular loop without leads, this flux affects the inter-dot hopping strength. Using perturbation theory in the inter-dot tunneling coupling $t$, where we treat $t$ as the small parameter compared to the on-site Coulomb repulsion $U$, we can derive the effective spin-exchange Hamiltonian from the quantum dot Hamiltonian $H_{S}$\cite{ref21,ref22} as

	\begin{equation}
		\begin{aligned}
			H_{\text {eff }}= & -t(1-n) \sum_{j k, \mu}\left(\hat{d}_{j \mu}^{\dagger} \hat{d}_{k \mu}+\text { H.c. }\right) \\
			& +J \sum_{j<k}\left(\hat{\boldsymbol{S}_{j}} \hat{\boldsymbol{S}_{k}}-\frac{1}{4} \hat{n}_{j} \hat{n}_{k}\right)+\chi \hat{\boldsymbol{S}_{1}}\left(\hat{\boldsymbol{S}_{2}} \times \hat{\boldsymbol{S}_{1}}\right),
		\end{aligned}
	\end{equation}

	where  $n$ is the the expectation value of the occupation number per dot, and $\hat{n}_{j}=\hat{d}_{j}^{\dagger} \hat{d}_{j}$ is the number operator for electrons on dot $j$. $\hat{\boldsymbol{S}_{1}}$, $\hat{\boldsymbol{S}_{2}}$ and $\hat{\boldsymbol{S}_{3}}$ are the spin operators on quantum dots, and $\mathbf{S}_{j}=\frac{1}{2} \sum_{\mu, \mu^{\prime}} \hat{d}_{j \mu}^{\dagger} \boldsymbol{\tau}_{\mu \mu^{\prime}} \hat{d}_{j \mu^{\prime}}$, $\boldsymbol{\tau}_{\mu \mu^{\prime}}$ are the Pauli matrices. The first term will vanish in the half-filling situation $(n=1)$. The second term is Heisenberg exchange interaction with $ J=   4 t^{2} / U$. The third term is the chiral term with chiral operator\cite{ref23} $ \widehat{S}_{1} \cdot\left(\widehat{S}_{2} \times \widehat{S}_{3}\right)$,  where $ \chi $ is the chiral interaction with  $\chi=24 t^{3} \sin \left(2 \pi \phi / \phi_{0}\right) / U^{2} $, and $ \phi $ is the magnetic flux enclosed by the TTQD structure. Here,  $\phi_{0}=h c / e $ is the unit of quantum flux. For simplicity, we let $ \varphi=2 \pi \phi / \phi_{0}$; thus, $ \chi=24 t^{3} \sin (\varphi) / U^{2}$.

	We can use Seeback coeffcient(S) to discribe thermopower,here is the defination  :
	\begin{equation}
		S \equiv-\left(\frac{V_{\mathrm{T}}}{\Delta T}\right)_{I=0}=\left(\frac{V}{\Delta T}\right)_{I=0} 
	\end{equation}.
	
	By employing the Hierarchical Equations of Motion (HEOM) methodology, we identify the bias voltage, denoted as $ V $, which neutralizes $ V_{\mathrm{T}} $. This allows for the accurate calculation of the Seebeck coefficient $S $, through strict adherence to its definition. In practical terms, it is beneficial to analyze the linear response regime because it simplifies the computational process. The expression for total current is given by
	\begin{equation}
		I=G V+L_{\mathrm{T}} \Delta T .
	\end{equation}
	
	Subsequently, the thermopower within this regime is determined through the formula
	
	\begin{equation}
		S=-\frac{L_{\mathrm{T}}}{G} ,
	\end{equation}
	
	where $L_{\mathrm{T}}$, which represents the thermoelectric coefficient, is defined as $\left(\frac{\partial I}{\partial \Delta T}\right)_{V=0}$. The conductance, $G$, is defined as $\left(\frac{\partial I}{\partial V}\right)_{\Delta T=0}$. Both parameters are essential for computing the responses of the electronic heat current or the voltage induced electron current when exposed to a minor temperature difference $\Delta T$ and a slight bias voltage $V$. Our study focuses on the low-temperature regime, where the contribution of phonons to the thermoelectric potential is strongly suppressed. The local excitations in quantum dots mainly arise from electron scattering processes and electron–electron correlation effects, while the influence of phonon vibrational modes on the electrons is very small and can therefore be neglected\cite{ref27}.

	The HEOM methodology explores the characteristics of quantum dots under both equilibrium and non-equilibrium conditions using the reduced density operator, which offers a universal framework applicable to any system's Hamiltonian. Here is a brief derivation. At time  $t$, let the reduced system density operator be $\rho(t)=\operatorname{tr}_{\text {res }} \rho_{\mathrm{T}}(t)$, where $\rho_{\mathrm{T}}(t)$ is the total density operator of the system plus reservoirs. $\rho(t)$ is related to the initial reduced system density operator at time $ t_{0} $ with the reduced Liouville-space propagator $ \mathcal{G}\left(t, t_{0}\right) $ 
	
	\begin{equation}
		\rho(t)=\mathcal{G}\left(t, t_{0}\right) \rho\left(t_{0}\right) .
	\end{equation}
	Based on Feynman-Vernon influence functional, the path-integral expression for the reduced Liouville-space propagator is
	
	\begin{equation}
		\mathcal{G}\left(\psi, t ; \psi_{0}, t_{0}\right)=\int_{\psi_{0}\left[t_{0}\right]}^{\psi[t]} \mathcal{D} \psi \mathrm{e}^{\mathrm{i} S[\psi]} \mathcal{F}[\psi] \mathrm{e}^{-\mathrm{i} S\left[\psi^{\prime}\right]} .
	\end{equation}

	The $S[\psi]$ is the classical action of the reduced system,  $\mathcal{F}[\psi]$ is the influence functional. According to Zhenhua Li et.'s study\cite{ref19}, using  Wick theorem and Grassmann algebra, we can ultimately derive the following formulation of $\mathcal{F}[\psi]$,
	
	\begin{equation}
		\mathcal{F}[\psi]=\exp \left\{-\int_{0}^{t} \mathrm{~d} \tau \mathcal{R}[\tau,\{\psi\}]\right\} 
	\end{equation}
	where  $\mathcal{R}[\tau,\{\psi\}]=\frac{i}{\hbar^{2}} \sum_{\alpha i \mu \sigma} \mathcal{A}_{i j \mu}^{\bar{\sigma}}[\psi(t)] \mathcal{B}_{{\alpha} i \mu}^{\sigma}[t, \psi]$, and $\sigma=+,-  , \bar{\sigma}=-\sigma$. Here,  $\mathcal{A}_{i s}^{\bar{\sigma}} $ and $ \mathcal{B}_{\alpha i \mu}^{\sigma} $ are the Grassmann variables defined as

	\begin{equation}
		\mathcal{A}_{i s}^{\bar{\sigma}}[\psi(t)]=d_{i s}^{\sigma}[\psi(t)]+d_{i s}^{\sigma}\left[\psi^{\prime}(t)\right]
	\end{equation}
	
	\begin{equation}
		\mathcal{B}_{\alpha i \mu}^{\sigma}[t, \psi]=-\mathrm{i}\left[B_{\alpha i \mu}^{ \sigma}(t, \psi)-B_{\alpha i \mu}^{\prime \sigma}\left(t, \psi^{\prime}\right)\right] ,
	\end{equation}

	with
	
	\begin{equation}
		\begin{array}{l}
			B_{\alpha i \mu}^{\sigma}(t, \psi)=\sum_{j} \int_{0}^{t} \mathrm{~d} \tau C_{\alpha i j \mu}^{\sigma}(t-\tau) d_{j \mu}^{\sigma}[\psi(\tau)], \\
			B_{\alpha i \mu}^{\prime \sigma}\left(t, \psi^{\prime}\right)=\sum_{j} \int_{0}^{t} \mathrm{~d} \tau C_{\alpha i j \mu}^{\bar{\sigma} *}(t-\tau) d_{j \mu}^{\sigma}\left[\psi^{\prime}(\tau)\right] ,
		\end{array}
	\end{equation}
	
	the $ C_{\alpha i j \mu}^{\sigma}(t)$ is reservoir correlation functions, in our calculation process, the expansion of $  C_{\alpha i j \mu}^{\sigma}(t)$ can be achieved through a series of exponential functions by implementing the fluctuation-dissipation theorem in conjunction with the Cauchy residue theorem and the Padé spectrum\cite{ref20} decomposition scheme of the Fermi function
	$C_{\alpha i j \mu}^{\sigma}(t)=\sum_{m=1}^{M} \eta_{\alpha i j \mu m}^{\sigma} \mathrm{e}^{-\gamma_{\alpha i j \mu m^{t}}^{\sigma} .}$
	
	The influence of bath enters the equations of motion with  M  exponentiations. The auxiliary density operators (ADOs)$  \left\{\rho_{j}^{n}=\rho_{j_{1} \ldots j_{n}}\right\}$  are determined by the time derivative of the influence functional, the index $ j \equiv(\sigma \mu m)$  corresponds to the transfer of an electron to/from  ($\sigma=+/-$)  the impurity state  $\mu$,$n=1,2,......,L$, denotes the truncated tier level, in this study, we choose $L=4$ is enough to get exact results. The final form can be reduced to the following compact form:
	
	\begin{equation}
		\begin{aligned}
			\dot{\rho}_{j_{1} \cdots j_{n}}^{(n)}= & -\left(\mathrm{i} \mathcal{L}+\sum_{r=1}^{n} \gamma_{j}\right) \rho_{j_{1} \cdots j_{n}}^{(n)}-\mathrm{i} \sum_{j} \mathcal{A}_{j} \rho_{j_{1} \cdots j_{n} j}^{(n+1)} \\
			& -\mathrm{i} \sum_{r=1}^{n}(-)^{n-r} \mathcal{C}_{j_{r}} \rho_{j_{1} \cdots j_{r-1} j_{r+1} \cdots j_{n}}^{(n-1)}.
		\end{aligned}
	\end{equation}
	
	$\mathcal{L}$ is the Liouvillian of dots, contains the $e-e$ interactions, and$\mathcal{L} \cdot \equiv\left[H_{\mathrm{dot}}, \cdot\right]$. The Grassmannian superoperators $\mathcal{A}_{\bar{j}} \equiv \mathcal{A}_{i s}^{\bar{\sigma}} \text { and } \mathcal{C}_{j} \equiv \mathcal{C}_{i j s m}^{\sigma}$ are defined via their fermionic actions on an operator $\hat{O}$  as $ \mathcal{A}_{j} \hat{O} \equiv\left[\hat{d}_{i s}^{\hat{\sigma}}, \hat{O}\right]$  and  $\mathcal{C}_{j} \hat{O} \equiv \eta_{j} \hat{d}_{i s}^{\sigma} \hat{O}+\eta_{j}^{*} \hat{O} \hat{d}_{i s}^{\sigma}$  respectively. 

Moreover, only first-tier auxiliary density operators is required to accurately calculate  the transient current through the
	electrode $\alpha$ .
	
	\begin{equation}
		I_{\alpha}(t)=\mathrm{i} \sum_{i \mu} \operatorname{tr}_{\mathrm{sys}}\left[\rho_{\alpha i \mu}^{\dagger}(t) \hat{d}_{i \mu}-\hat{d}_{i \mu}^{\dagger} \rho_{\alpha i \mu}^{-}(t)\right] .
	\end{equation}

In addition, We obtain spectral function by calculating the time evolution of the correlation functions $C_{a_{i \mu} a_{i \mu}^{\dagger}}(t)$ and $C_{a_{i \mu}^{\dagger} a_{i \mu}}(t)$, and by performing a semi-Fourier transform on them, the spectral function $A_{i \mu}(\omega)$ of the $\mu$ spin in the ith quantum dot can be directly obtained. We can obtain the spectral density function of quantum dots, and the correlation function can be obtained from the spectral density function by using the fluctuation-dissipation theorem.

\begin{equation}
\begin{array}{c}
A_{i \mu}(\omega)=\frac{1}{\pi} \operatorname{Re}\left\{\int_{0}^{\infty} d t\left\{C_{a_{i \mu} a_{i \mu}^{\dagger}}(t)+\left[C_{a_{i \mu}^{\dagger} a_{i \mu}}(t)\right]^{*}\right\} e^{-i \omega t}\right\}
\end{array}
\end{equation}

\section{Results and Discussion}

We first calculated the $ZT$ value and thermoelectric potential at low temperature conditions of $k_B T=0.04meV$, as the intra-dot transition strength $ t_{13} $ varies, as shown in figure 1(b). And we set $t_{12}=t_{23}=0.4meV$, each QD's on-site energy is $\varepsilon=-U/2=-0.5meV$. As $t_{13}$ increases from 0 to 1 $meV$, it can be seen that the maximum $ZT$ can reach 4.46, when $t_{13}=0.55meV$. The trend of themopower is very similar to $ZT$, The themopower $ S $ changes more rapidly, reaching an extremum at $t_{13}=0.52$, and the variation curve is relatively smoother. During this variation process, our calculation results show that the changes in thermal conductance $\kappa$ and electrical conductance $ G $ are relatively synchronous near the  extremum of $ZT$, because $G/ \kappa $ remains stable. Therefore, we conclude that the main factor affecting the thermoelectric figure of merit in this process is the thermopower.
\begin{figure}[ht] 
	\centering 
	\includegraphics[width=0.4\textwidth]{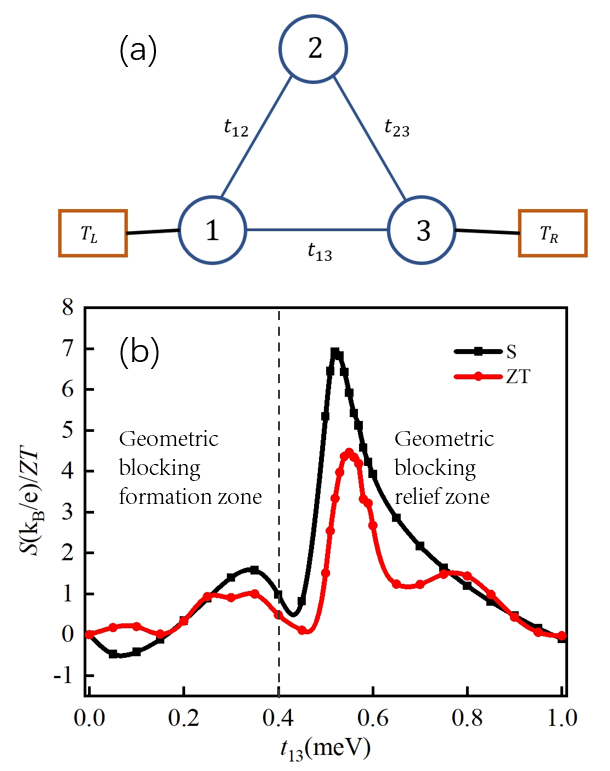} 
	\caption{(a)Quantum dot 1 and quantum dot 3 are connected to the left and right electrodes.The coupling strength between each electrode and the quantum dot is 0.2 meV.(b)shows how the thermoelectric parameters change with the intradot tunneling strength  $t_{13}$
. The black line indicates the Seebeck coefficient, while the red line indicates the  $ZT$
 value. The tunneling strength  $t_{13}$ranges from 0 to 1 \, \text{meV}. The other parameters are set as follows:  $U = 1 \, \text{meV}$, $\epsilon = -0.5 \, \text{meV}$,  $t_{12} = t_{23} = 0.4 \, \text{meV}$, and  $k_B T = 0.04 \, \text{meV}$} 
	\label{fig:example-image} 
\end{figure}

To further investigate the origin of the extreme value of thermopower, we calculate the electric current values in two scenarios: one where the electrodes had only a bias voltage and another where they had only a temperature difference,as shown in figure2. Voltage induced electron current is the current that results when only voltage difference is applied across the two leads, while electronic heat current is the current measured when only small temperature difference is applied across the two leads. The voltage difference and temperature difference between the left and right electrodes were controlled on the same energy scale. Initially, when $t_{13}$ increased, both currents reached a peak value before decreasing to near zero. The initial rise in current results from the opening of a new transport channel between quantum dot 1 and quantum dot 3 as $t_{13}$ incrises. Interestingly, at low temperatures, the electronic heat current exhibited some positive values at the beginning of the $t_{13}$ increase due to the excited carriers are holes. As $t_{13}$ increases to approach the values of $t_{12}$ and $t_{23}$, the system gradually shifts from a linear triple quantum dots arrangement to a triangular configuration. This geometric transition plays a key role in the development of geometric blockade effects. In the triangular case, geometric phase become more pronounced, resulting in destructive interference that restricts electron transport across the device. This change is evident in both the voltage induced electron current and electronic heat current, both of which show a significant decrease as the structure approaches a triangular form. The stronger geometric blockade leads to additional phase cancellations due to the increased number of transport pathways, further suppressing transport. These results underscore how the interplay between tunneling couplings and device geometry critically influences both charge and energy transport in nanoscale quantum dots systems.

\begin{figure}[ht] 
	\centering 

	\includegraphics[width=0.4\textwidth]{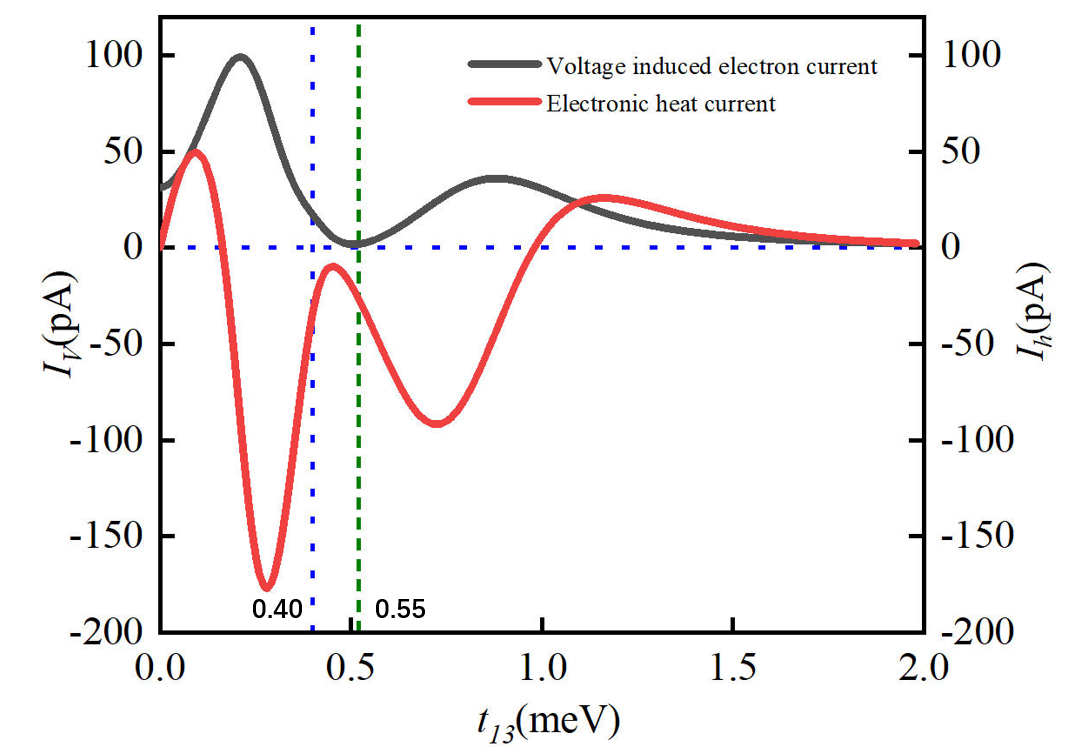}
	\caption{shows the relationship between the electronic heat current and the voltage induced electron current as $t_{13}$ varies. The red line represents the current under a temperature difference between the electrodes, while the black line represents the current under a voltage difference. Other parameters are:  $k_B T = 0.04 \, \text{meV}$,  $\Delta T = 0.004 \, \text{meV}$,  $\Delta V = 0.004 \, \text{meV}$;  $t_{12} = t_{23} = 0.4 \, \text{meV}$,  $U = 1 \, \text{meV}$,  $\epsilon = -0.5 \, \text{meV}$} 
	\label{fig:example-image} 
\end{figure}

As $t_{13}$ continued to increase, geometric blockade is breaked. Notably, the extreme value of thermoelectric potential did not coincide with the maximum blockade but occurred during the process of blockade removal. Our calculations revealed that during blockade removal, the thermoelectric current recovered faster and exhibited larger values compared to the conduction current. This finding suggests that the thermoelectric current induced by the temperature difference is more sensitive to the removal of the blockade than the conduction current caused by the bias voltage,which leads to the extremum of themopower. As a result, the maximum thermopower is observed when the system strikes a special state—neither fully blocked nor completely delocalized—achieving optimal energy conversion efficiency. These results showcase how subtle shifts in coupling and geometry control transport properties in quantum dots systems. The pronounced dependence of thermoelectric performance on the lifting of the geometric blockade offers practical insights for engineering high-performance nanoscale thermoelectric devices.

In quantum transport theory, a widely used method for calculating $G$ and $L_{T}$ conductance is the Landauer formula. Starting from this formula, we can derive the following expression for the thermoelectric potential in equilibrium. 

\begin{equation}
		\begin{aligned}
			S_{\text {Landaucer }}=-\frac{1}{e T} \frac{\int d \omega(\omega-\mu) f^{\prime}(\omega) \mathcal{T}(\omega)}{\int d \omega f^{\prime}(\omega) \mathcal{T}(\omega)},
		\end{aligned}
	\end{equation}
where $f^{\prime}(\omega)$ is the derivative of the Fermi-Dirac distribution, $ \mathcal{T}(\omega)$ is frequency resolved transmission probability.

If we assume a very large bandwidth, corresponding to the wide-band approximation, the transmission function $ \mathcal{T}(\omega)$ becomes proportional to the total spectral function of the system $A_{T}(\omega)$, differing only by an overall proportionality factor involving the constant coupling strengths, we can deduce: 
\begin{equation}
		\begin{aligned}
			S_{\text {Landaucer }}=-\frac{1}{e T} \frac{\int d \omega(\omega-\mu) f^{\prime}(\omega) A_{T}(\omega)}{\int d \omega f^{\prime}(\omega) A_{T}(\omega)},
		\end{aligned}
	\end{equation}

We calculated the local spectral function of dot1 during the variation of $t_{13}$ under temperature difference conditions, as shown in the figure3.
It can be observed that initially, as $t_{13}$ increases, the geometric blockade gradually forms, and the electrons originally near zero energy begin to shift towards a new state. This is reflected in the local spectral function as the peak near zero energy and another peak approaching each other. When the blockade is relatively large, they merge into a single peak, forming a new state. With further increases in $t_{13}$ and as the geometric blockade is lifted, this merged peak splits again, restoring the two separate resonances seen at lower couplings.   
We calculated the full width at half maximum (FWHM) of the splitting and merging peaks as$t_{13}$ varies and found that it changes linearly with the increase of $t_{13}$.
\begin{figure}[ht] 
	\centering 
	\includegraphics[width=0.5\textwidth]{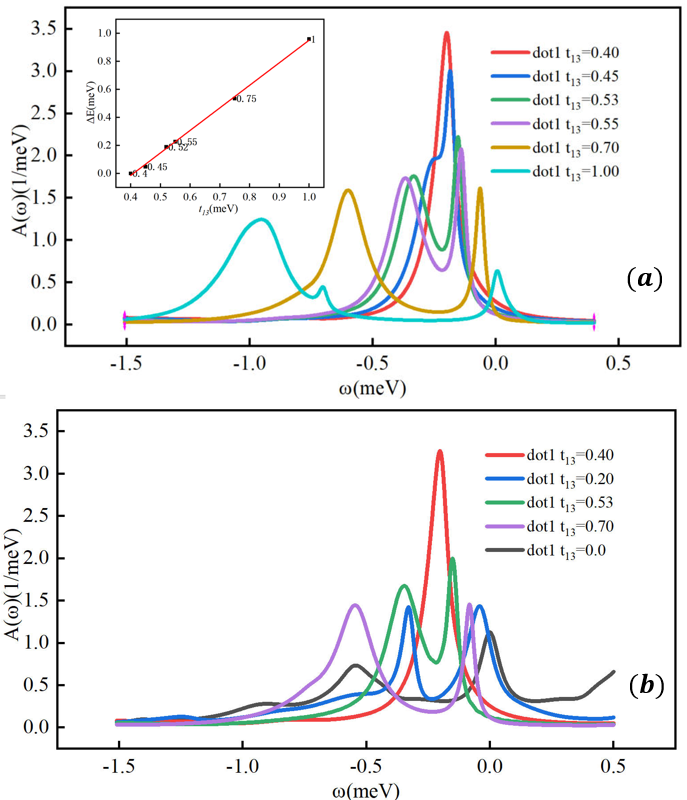} 
	\caption{(a) Local spectrum function of quantum dot 1 under different coupling strengths between quantum dots 1 and 3, in the presence of a temperature difference between the left and right electrodes. The inset shows the variation of the horizontal coordinate difference between the split peaks with respect to  $t_{13}$.(b) Local spectrum function of quantum dot 1 in the presence of a potential difference between the left and right electrodes.} 
	\label{fig:example-image} 
\end{figure}

Analysis of the local spectral function also sheds light on the different responses of the electronic heat current and voltage induced electron current.  During the lifting of the blockade, the structure of the spectral function changes rapidly—most notably, one of the split peaks approaches the Fermi level.  The  electronic heat current is calculated from the difference between the spectral weight above and below the Fermi surface, which makes it especially sensitive to excitations across a range of energies.  In contrast, the voltage induced electron current mainly depends on the spectral weight near the Fermi energy, which reflects only those states that participate directly in transport under a bias. By directly comparing their spectral signatures, it becomes clear that the electronic heat current responds much more strongly to changes in t13 than the voltage induced electron current does.  This greater sensitivity results in a pronounced peak in the  electronic heat current potential, occurring precisely at the stage where the  electronic heat current changes most rapidly as the blockade is removed.  Accordingly, the figure of merit (ZT) reaches its maximum in this intermediate regime, highlighting the critical influence of energy-selective transport and geometric configuration on thermoelectric efficiency.

To further analyze, we calculated the numerator and denominator parts of Eq.(19) qualitatively, we calculated the local spectral function of all dots. The denominator and numerator in the Landauer formula exhibit fundamentally different responses to this spectral evolution.(i) Denominator (conductance term):$\int d \omega f^{\prime}(\omega) A_{T}(\omega)$ increases monotonically with $t_{13}$, it grows continuously due to the broadening of transmission channels. This reflects the conventional wisdom that stronger coupling enhances conductance. (ii) Numerator (thermoelectric term):$\int d \omega(\omega-\mu) f^{\prime}(\omega) A_{T}(\omega)$ exhibits non-monotonic behavior, reaching a pronounced maximum during the intermediate splitting regime when $t_{13}=0.55meV$.
At low temperatures, we use the Sommerfeld expansion, from which the denominator is found to be proportional to $\frac{\partial A_{T}\left(\omega, t_{13}\right)}{\partial \omega}|_{\omega=\mu}$. This implies the energy gradient achieves its maximum value precisely when the spectral peak begins to split but has not yet fully separated—a transient regime where the spectral function exhibits the steepest energy-dependent variation, which further explains why we obtain the peak in the thermoelectric potential.

We further investigate the temperature dependence of this phenomenon, we keep other parameters unchanged, set $t_{13}=0.55meV$, as illustrated in Fig.4(a) and 4(b), which present the calculated thermopower and figure of merit $ZT$, respectively, at elevated temperatures.  It is evident that both the thermopower and $ZT$ decrease rapidly with increasing temperature. This degradation originates from several temperature-induced effects. First, the thermal broadening of the Fermi-Dirac distribution function becomes significant at elevated temperatures, electrons over a wider energy range contribute to transport, weakening the energy filtering mechanism that is crucial for large thermopower. Second, inelastic scattering processes, including electron-phonon interactions, become more pronounced at higher temperatures, reducing the phase coherence of electron transport through the quantum dot. This leads to a degradation of the quantum interference effects that contribute to the thermoelectric response. These combined effects explain the temperature sensitivity of the thermoelectric performance in the our system.

\begin{figure}[ht] 
	\centering 
	\includegraphics[width=0.5\textwidth]{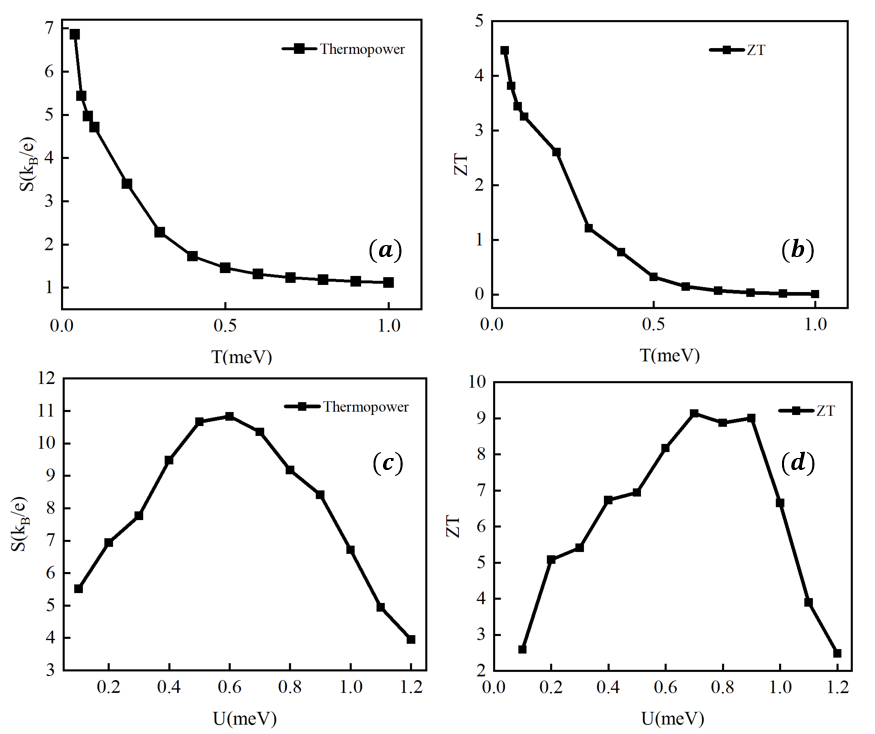} 
	\caption{(a) and (b) show how the thermoelectric parameters change with the T, (c) and (d) show how the thermoelectric parameters change with the U} 
	\label{fig:example-image} 
\end{figure}

Fig. 4(c) and 4(d) display the calculated thermopower and figure of merit $ZT$ as functions of the on-site Coulomb interaction U, we also keep other parameters unchanged, set $t_{13}=0.55meV$, while maintaining particle-hole symmetry.  A non-monotonic behavior is observed: both quantities initially increase and then decrease as U increases from weak to strong coupling regime. This intriguing trend can be understood as follows.  In the weak coupling regime, the enhancement of Coulomb interaction strengthens the many-body correlations and facilitates the formation of the Kondo resonance, which enhances the thermoelectric response near the Fermi level. However, as U further increases into the strong coupling regime, the on-site Coulomb repulsion becomes dominant, making it energetically unfavorable for double occupancy of the quantum dot.  This Coulomb blockade effect significantly suppresses electron tunneling between the dot and the electrodes, resulting in reduced conductance and degraded thermoelectric performance.  The existence of an optimal U value indicates that thermoelectric efficiency is maximized when many-body correlations and charge fluctuations are appropriately balanced. This non-monotonic dependence on U underscores the intrinsic nature of this thermoelectric phenomenon as a manifestation of strongly correlated electron physics, where the interplay between Coulomb interaction and quantum fluctuations plays a decisive role.

In summery, these observations illustrate not only uncovers the intricate physics of triple quantum dot systems but also establishes clear links between energy level evolution and the contrasting behaviors of charge and heat transport.  Such insights are valuable for guiding the design and optimization of high-performance quantum-dot-based thermoelectric devices.

\section{Acknowledgements}  The support from the Natural Science Foundation of China (Grant Nos. 12274454, 11774418, 11374363, 11674317, 11974348, 11834014 and 21373191), the Strategic Priority Research Program of CAS (Grants No. XDB28000000 and No. XDB33000000), the Training Program of Major Research plan of NSFC (Grant No. 92165105).

\section{Data Availability}
The data that support the findings of this study are available from the corresponding author upon reasonable request.


\begin{thebibliography}{99}

     \bibitem{ref1} Bell, L. E. (2008). Science, 321(5895), 1457–1461. 
	\bibitem{ref2} Franz R and Wiedemann D 1853 Ann. Phys. 165 497
	\bibitem{ref3} N.W. Ashcroft, N.D. Mermin, Solid-State Physics, Saunders College Publishing,
	Philadelphia, 1967.
	\bibitem{ref4} B. Kubala, J. König, J. Pekola, Phys. Rev. Lett. 100 (2008) 066801.
	\bibitem{ref5} P. Murphy, S. Mukerjee, J. Moore, Phys. Rev. B 78 (2008) 161406. 
	\bibitem{ref6} A. Majumdar, Science 303 (2004) 777.
	\bibitem{ref7} T.C. Harman, P.J. Taylor, M.P. Walsh, B.E. LaForge, Science 297 (2002) 2229.
	\bibitem{ref8} A.A. Balandin, O.L. Lazarenkova, Appl. Phys. Lett. 82 (2003) 415.
	\bibitem{ref9} P. Reddy, S.Y. Jang, R.A. Segalman, A. Majumdar, Science 315 (2007) 1568.
	\bibitem{ref10} R. Scheibner, M. Konig, D. Reuter, A.D. Wieck, H. Buhmann, L.W. Molenkamp,
	New J. Phys. 10 (2008) 083016.
	\bibitem{ref11} S.F. Svensson, A.I. Persson, E.A. Hoffmann, N. Nakpathomkun, H.A. Nilsson, H.Q.Xu, L. Samuelson, H. Linke, New J. Phys. 14 (2012) 033041.
	
	\bibitem{ref12}	H. Thierschmann, M. Henke, J. Knorr, L. Maier, C. Heyn, W. Hansen, H. Buhmann, L.W. Molenkamp, New J. Phys. 15 (2013) 123010
	\bibitem{ref13}  M. Wierzbicki, R. Swirkowicz, Phys. Rev. B 84 (2011) 075410.
	\bibitem{ref14}Q. Wang, Q.H. Xie, Y.H. Nie, W. Ren, Phys. Rev. B 87 (2013) 075102.
	\bibitem{ref15}X. Lu, J.-S. Wang, W.G. Morrel, X. Ni, C.-Q. Wu, B. Li, J. Phys. Condens. Matter 27 (2015) 035301.
	\bibitem{ref16}  Weymann I, BuŁka B R and Barnas J ´ 2011 Phys. Rev. B 83 195302
	\bibitem{ref17}Niklas M, Trottmann A, Donarini A and Grifoni M 2017 Phys. Rev. B
	95 115133 
	\bibitem{ref18} Y. D. Wang and Z. G. Zhu and J. H. Wei and Y. J. Yan  EPL (Europhysics Letters), 2020, 130(1): 17003.
	\bibitem{ref19}	Li Z H, Tong N H, Zheng X, Hou D, Wei J H, Hu J and Yan Y J 2012
	Phys. Rev. Lett. 109 266403
	\bibitem{ref20} Ye L Z, Wang X L, Hou D, Xu R X, Zheng X and Yan Y J 2016 WIREs
	Comput. Mol. Sci. 6 608
	\bibitem{ref21}C. W. J. Beenakker and A. A. M. Staring, Phys. Rev. B 46, 9667–9676 (1992).
	\bibitem{ref22} Scarola V W and Das Sarma S 2005 Phys. Rev. A 71 032340
	\bibitem{ref23} Wen X G, Wilczek F and Zee A 1989 Phys. Rev. B 39 11413
	\bibitem{ref24}N. Byers and C. Yang, Physical review letters 7, 46 (1961).
     \bibitem{ref25}T. E. Humphrey, R. Newbury, R. P. Taylor, and H. Linke,  Phys. Rev. Lett. 89, 116801 (2002).
     \bibitem{ref26}R. Scheibner, H. Buhmann, D. Reuter, A. D. Wieck, and L. W. Molenkamp,  Phys. Rev. Lett. 95, 176602 (2005).
     \bibitem{ref27}Mani P, Nakpathomkun N, Hoffmann E A, et al. Nano Lett., 2011, 11(11):4679–4681.
     \bibitem{ref28}M. Josefsson, A. Svilans, A. M. Burke, E. A. Hoffmann, S. Fahlvik, C. Thelander, M. Leijnse, and H. Linke, Nature Nanotechnology 13,    920(2018).
     \bibitem{ref29} Y. Tanimura, J. Chem. Phys. 153, 020901 (2020).
     \bibitem{ref30}Gomez-Silva, G. Orellana, P. A., Anda, E. V., J. Appl. Phys. 123, 085706 (2018).
     \bibitem{ref31}Pirot, B. R., Abdullah, N. R., Ari Karim Ahmed, Physica B: Condensed Matter 629, 413646 (2022a).
\end{thebibliography}
\end{document}